\documentclass[sigconf]{acmart}
\AtBeginDocument{%
  }

\renewcommand\footnotetextcopyrightpermission[1]{}
\setcopyright{none}

\usepackage{graphicx} 
\usepackage{booktabs} 
\usepackage{threeparttable} 
\usepackage{xcolor} 
\usepackage[skins,breakable]{tcolorbox}
\usepackage[utf8]{inputenc}
\usepackage{amsmath}
\usepackage{enumitem} 
\usepackage{adjustbox}
\usepackage{booktabs}
\usepackage{tabularx}
\usepackage{caption}

\NewDocumentCommand{\findingbox}{ m m }{%
    \begin{tcolorbox}[
        enhanced,
        colback=gray!10,
        frame hidden,
        boxrule=0pt,
        arc=2mm,
        boxsep=2pt,
        left=6pt,
        right=6pt,
        top=6pt,
        bottom=2pt,
        title=\textbf{#1}, 
        coltitle=black,
        fonttitle=\bfseries\sffamily,
        attach boxed title to top left={yshift=-2mm, xshift=3mm},
        boxed title style={colback=white,boxrule=0pt,frame hidden,colback=gray!40}
    ]
    #2 
    \end{tcolorbox}
}

\usepackage{pifont} 

\newcommand{\parabf}[1]{\noindent\textbf{#1}}

\newcommand{\bench}{E2EDevBench}
\newcommand{\agentd}{\texttt{SDAgent-Single}}
\newcommand{\agentdt}{\texttt{SDAgent-DT}}
\newcommand{\agentddt}{\texttt{SDAgent-DDT}}

\begin{document}

\title{Benchmarking and Studying the LLM-based Agent System in End-to-End Software Development}


\author{Zhengran Zeng}
\authornote{Both authors contributed equally to this research.}
\email{zhengranzeng@stu.pku.edu.cn} 
\affiliation{%
  \institution{Peking University}
  \city{Beijing}
  \country{China}
}
\author{Yixin Li}
\authornotemark[1] 
\email{leason_lyx@stu.pku.edu.cn} 
\affiliation{%
  \institution{Peking University}
  \city{Beijing}
  \country{China}
}
\author{Rui Xie}
\email{ruixie@pku.edu.cn} 
\affiliation{%
  \institution{Peking University}
  \city{Beijing}
  \country{China}
}
\author{Wei Ye}
\email{wye@pku.edu.cn} 
\affiliation{%
  \institution{Peking University}
  \city{Beijing}
  \country{China}
}
\author{Shikun Zhang}
\email{zhangsk@pku.edu.cn} 
\affiliation{%
  \institution{Peking University}
  \city{Beijing}
  \country{China}
}

\renewcommand{\shortauthors}{Trovato et al.}

\begin{abstract}
The development of LLM-based autonomous agents for end-to-end software development represents a significant paradigm shift in software engineering. However, the scientific evaluation of these systems is hampered by significant challenges, including overly simplistic benchmarks and the difficulty of conducting fair comparisons between different agent architectures due to confounding implementation variables. To address these limitations, we first construct a challenging and dynamically curated \bench{} to simulate realistic development scenarios. Second, we propose a hybrid evaluation framework that combines test-case-based functional assessment with fine-grained, LLM-based requirement verification. Using this framework, we conduct a controlled empirical study on three representative agent architectures implemented upon a unified foundation to isolate the impact of workflow design. Our findings reveal that state-of-the-art agents can fulfill approximately 50\% of requirements on \bench{}, but their success is critically dependent on the architectural strategy for task decomposition and collaboration. Furthermore, our analysis indicates that the primary bottleneck is the omission of requirements and inadequate self-verification. This work provides the community with a more realistic benchmark, a comprehensive evaluation framework, and crucial insights into the current capabilities and core challenges of software development agents, guiding future research toward enhancing requirement comprehension and planning.
\end{abstract}

\maketitle

\section{Introduction}
The rapid advancement of Large Language Models (LLMs) in recent years has driven significant progress in the development of LLM-based autonomous agents~\cite{surveyonllmbasedautonomousagents}. These agents are poised to introduce a paradigm shift within the field of software engineering, having already demonstrated capabilities across a range of tasks such as issue fixing, feature implementation, and test generation~\cite{SurveyofBenchmarks,AgentsforSoftwareEngineeringSurvey1,AiderDocumentation,SWEagent,RepairBench,SoftwareTestingSurvey}. 
Among these applications, the automation of end-to-end software development, which spans the entire lifecycle from initial requirement specification to the delivery of functional software, has emerged as a particularly prominent and critical area of research~\cite{AgentsforSoftwareEngineeringSurvey1}.

With the goal of automating end-to-end software development, the research community has proposed a variety of agent systems, including ChatDev and MetaGPT~\cite{ChatDev,MetaGPT}. To measure their performance, corresponding benchmarks like SoftwareDev and Commit0 have also emerged\cite{MetaGPT,commit0}. However, despite this growing body of research, the scientific and comprehensive evaluation of these agent systems remains a significant challenge. Specifically, current evaluation approaches often fall short in two significant ways. Firstly, they may present tasks that are overly simplified and lack real-world complexity, a limitation observable in benchmarks such as SoftwareDev~\cite{MetaGPT}. Secondly, in the pursuit of automated and quantitative comparisons, frameworks like Commit0~\cite{commit0} provide agents with exhaustive test suites or detailed code specifications. Such a setup substantially differs from authentic development workflows, thereby limiting our ability to gauge an agent's true capabilities in a realistic setting.

Moreover, systematic comparisons between different agent architectures, such as single-agent and multi-agent systems or different workflows, remain limited. A major challenge for fair comparison is the significant variation in how these agents are implemented. Differences in their underlying toolsets and engineering optimizations create confounding variables, making it difficult to perform a fair comparison and analysis. Consequently, a performance gain might be attributed not to a superior architectural design, but rather to better engineering or more powerful tools. Additionally, most existing studies evaluate these agents on relatively simple tasks~\cite{MetaGPT,ChatDev,AgileCoder}. This raises a critical question: can their findings be generalized to the more complex and realistic scenarios found in real-world software development?

To address the aforementioned challenges, this study aims to establish a more robust foundation for the evaluation and analysis of agents for end-to-end software development. First, we construct and introduce a more challenging software development dataset, dynamically curated from the latest PyPI projects~\cite{PyPI}. This approach mitigates the risk of data leakage and simulates realistic, demanding development scenarios.
Second, to meet the complex evaluation demands of end-to-end tasks, we propose a hybrid evaluation methodology. This methodology integrates test-case-based functional assessment with a fine-grained, LLM-based evaluation of requirement conformance, enabling a more precise measurement of an agent's performance in fulfilling user needs. Building on this, to facilitate fair comparisons and investigate the differences between architectures, we implement three representative agent systems based on a unified engineering standard and toolset. This controlled experimental environment allows us to isolate and study the impact of agent architecture and workflow design on performance.

Through our empirical study, we aim to address several core research questions. Specifically, we seek to: (1) assess the capability of current software development agents in completing realistic end-to-end software development tasks; (2) analyze the impact of different agent workflows (e.g., single-agent versus multi-agent) on performance; and (3) determine the primary causes of agent failure during the development process.

Our research yields several key insights. First, we find that state-of-the-art agents, when powered by the most advanced LLMs, can successfully fulfill approximately 50\% of the requirements in our challenging benchmark. Meanwhile, their performance is highly dependent on their architectural design and orchestration strategy. Specifically, effective task decomposition and multi-agent collaboration significantly reduce problem complexity, while poor architectural choices can severely hinder performance. Furthermore, our analysis reveals that the primary reason for failure is not incorrect code implementation, but rather the omission of requirements and inadequate self-verification. This suggests that the core bottleneck lies in the upstream stages of planning and understanding, highlighting the need for further research into requirement comprehension and validation.

The primary contributions of this research can be summarized as follows:
\begin{enumerate}[label=\arabic*., topsep=0.5em, itemsep=0.2em]
    \item \textbf{Benchmark:} A more challenging and dynamically updated benchmark dataset for end-to-end software development.
    \item \textbf{Evaluation Framework:} A comprehensive evaluation framework designed for assessing end-to-end development tasks by combining automated test execution with fine-grained requirement verification.
    \item \textbf{Approach:} A standardized, open-source implementation of end-to-end development agent system, providing a fair baseline for future comparative research.
    \item \textbf{Study:} An in-depth empirical study that reveals the capability boundaries and critical bottlenecks of current software development agents.
\end{enumerate}

\section{Background and Related Work}

\subsection{End-to-End (E2E) Software Development}

End-to-end software development aims to automate the entire Software Development Life Cycle (SDLC)~\cite{AgentsforSoftwareEngineeringSurvey1}, from initial requirement specification to final product delivery, through a single, autonomous system. The typical task paradigm involves an agent receiving a software requirement specified in natural language (e.g., ``develop a 2048 game'') and autonomously generating a complete, functional, and executable code repository.

The recent advancement of LLMs has significantly accelerated the automation of software engineering tasks. For instance, a variety of LLM-driven tools have proven effective for individual stages of the Software Development Life Cycle (SDLC)~\cite{llmforSoftwareEngineeringreview}, such as requirements engineering~\cite{RequirementsEngineeringSurvey}, function-level code generation~\cite{HumanEval,MBPP}, and software testing~\cite{SoftwareTestingSurvey}. Building on these successes, the end-to-end development paradigm aims to integrate these separate steps into a single, seamless pipeline---from initial requirement analysis to final deployment. Given its great potential and inherent challenges, end-to-end software development is now widely seen as a critical frontier in automation research.

\subsection{E2E Software Development Benchmarks}
\label{sec:benchmark_background}
To facilitate the evaluation of automated systems on end-to-end tasks, the research community has proposed several benchmarks. Unlike earlier benchmarks focused on function- or file-level code generation (e.g., HumanEval~\cite{HumanEval}, MBPP~\cite{MBPP}, APPS~\cite{APPS}, and LiveCodeBench~\cite{LiveCodeBench}), benchmarks for end-to-end development place a greater emphasis on simulating complex, realistic development scenarios.

For instance, the SoftwareDev~\cite{MetaGPT} benchmark features 70 diverse tasks, from arcade games to business systems, that require agents to manage multi-file dependencies and project-level architecture. Similarly, DevEval~\cite{DevEval(b)} uses real projects from GitHub to create its tasks, and it evaluates agents across five distinct stages: software design, environment setup, implementation, acceptance testing, and unit testing. ProjectEval~\cite{ProjectEval}, which also draws data from SoftwareDev, measures the project-level code generation capabilities of LLMs by providing inputs at different abstraction levels (i.e., from requirements document to code skeleton) and using a predefined test suite for evaluation. In a different approach, Commit0~\cite{commit0} presents agents with a code skeleton and a requirements document from real GitHub projects, tasking them with completing the code to pass the original tests.

However, existing benchmarks exhibit notable limitations:
\begin{enumerate}[label=\arabic*., topsep=0.5em, itemsep=0.2em, leftmargin=*]
    \item \textbf{Insufficient Challenge:} Many benchmarks, such as SoftwareDev~\cite{MetaGPT}, contain tasks that remain at the level of ``toy projects.'' Meanwhile, the reference implementations for DevEval~\cite{DevEval(b)} and ProjectEval~\cite{ProjectEval} are often modest in size (typically 300-400 lines of code), indicating a complexity gap compared to industrial-grade projects.

    \item \textbf{Limited Evaluation Metrics:} Many studies (e.g., ChatDev~\cite{ChatDev}) rely on metrics that are often superficial. For instance, ``completeness'' is typically assessed by checking for code placeholders like ``TODO''~\cite{ChatDev}, while ``consistency'' is often measured by vector similarity between requirements and code~\cite{CodeS,ChatDev,YABLoCo}. Similarly, ``executability'' is often just a coarse-grained check to ensure the code runs without crashing~\cite{MetaGPT,AgileCoder,ChatDev,DevEval(a)}. These indirect metrics fail to provide an accurate reflection of software quality.
    
    \item \textbf{Compromised Realism in Evaluation:} To resolve interface mismatches between agent-generated code and predefined test cases, benchmarks like DevEval~\cite{DevEval(b)}, ProjectEval~\cite{ProjectEval}, and Commit0~\cite{commit0} provide test cases or even code skeleton designs as input to the agent. While this approach simplifies evaluation, it deviates significantly from real-world scenarios where only a requirements document is provided.
\end{enumerate}
In light of these shortcomings, our research is dedicated to constructing a benchmark that is both more challenging and more closely aligned with real-world evaluation methodologies. We extract tasks from high-quality PyPI projects and design an evaluation framework that combines automated test migration with objective requirement verification, aiming to measure the true capabilities of agents more equitably. The specific details of our approach are presented in Section~\ref{sec:benchmark}.

\subsection{E2E Software Development Agents}
\label{sec:agent_background}
Software Development Agents are autonomous systems based on LLMs that are capable of performing software engineering tasks in a self-directed manner. A typical agent system comprises four core components: Planning, Memory, Perception, and Action~\cite{AgentsforSoftwareEngineeringSurvey1,LLMBasedAgentsSurvey,surveyonllmbasedautonomousagents,LLMstoLLM-basedAgentsforSE}. The planning component is responsible for decomposing high-level objectives into executable sub-tasks. The memory component is utilized for storing and retrieving historical information. The perception component is responsible for understanding user instructions and the state of the environment. The action component interacts with the environment by invoking external tools.

Currently, agent architectures applied to end-to-end software development can be broadly categorized into two main classes:

\begin{enumerate}[leftmargin=*]
\item \textbf{Single-Agent:} Exemplified by systems like SWE-Agent~\cite{SWEagent}, the core idea of this approach is to enhance the perception and action capabilities of an individual agent. For instance, SWE-Agent significantly improves an agent's ability to comprehend and edit complex code by designing a specialized toolset for interacting with a codebase. Similar works, such as ReAct~\cite{ReAct}, CodeAgent~\cite{CodeAgent} and OpenHands~\cite{OpenHands}, also concentrate on optimizing tool definitions and environmental perception to improve the task execution capabilities of a single agent.

\item \textbf{Multi-Agent:} Represented by systems like ChatDev~\cite{ChatDev}, AgentVerse~\cite{AgentVerse} and MetaGPT~\cite{MetaGPT}, the research focus of this approach lies in task decomposition and workflow orchestration. By simulating collaborative software development models, such as waterfall\cite{waterfall} and agile development~\cite{agilemodel}, these systems break down a complex task into multiple sub-tasks. These sub-tasks are then assigned to agents with distinct roles (e.g., ``Designer'',`` Programmer'', and ``Tester'') for collaborative completion.
\end{enumerate}

Although researchers have proposed various agent architectures, a systematic comparative analysis of these different approaches remains insufficient. A primary challenge is that the performance of agent systems is heavily influenced by their engineering implementation. Research on single-agent systems tends to focus on the perception and action components~\cite{SWEagent,OpenHands}; consequently, their engineering implementations and toolsets are often more sophisticated. In contrast, the focus of multi-agent systems is on workflow orchestration~\cite{ChatDev,MetaGPT}, and their underlying tools may be comparatively simpler. This disparity makes it difficult to conduct direct comparisons that yield reliable conclusions about the inherent merits or demerits of the architectures themselves.

Furthermore, as previously noted, most existing research has been conducted on relatively simple datasets. For example, some studies on multi-agent systems that explore different workflows have compared various orchestration methods under unified experimental conditions~\cite{ChatDev,MetaGPT}. However, the generalizability of their conclusions is limited by the simplicity of the datasets used (Humaneval~\cite{HumanEval} and MBPP~\cite{MBPP}). 
Therefore, there is a pressing need for research that conducts a fair and in-depth comparison of different agent architectures in more challenging, realistic scenarios, based on unified engineering standards. To address this gap, we implement and evaluate multiple representative agent systems on a carefully designed benchmark within a controlled experimental environment, enabling us to investigate the intrinsic impact of architectural and workflow design on performance.

\section{Study Design}

\subsection{Benchmark}
\label{sec:benchmark}
In order to accurately assess the performance of agents in realistic end-to-end software development scenarios, a more suitable benchmark is required to address the limitations of existing ones. As described in Section~\ref{sec:benchmark_background}, current benchmarks often fall into two categories. Firstly, datasets like SoftwareDev~\cite{MetaGPT} and SRDD~\cite{ChatDev} feature tasks of limited complexity, such as implementing small-scale applications or games, which do not adequately represent real-world engineering challenges. Secondly, while more challenging benchmarks like Commit0~\cite{commit0}, DevEval~\cite{DevEval(b)}, HumanEvo~\cite{HumanEvo} exist, they often require supplementary artifacts such as pre-defined project designs or test cases. This requirement deviates from authentic development workflows, where agents typically operate from high-level requirements alone.
 
\begin{figure*}[h]
    \centering
    \includegraphics[width=1\textwidth]{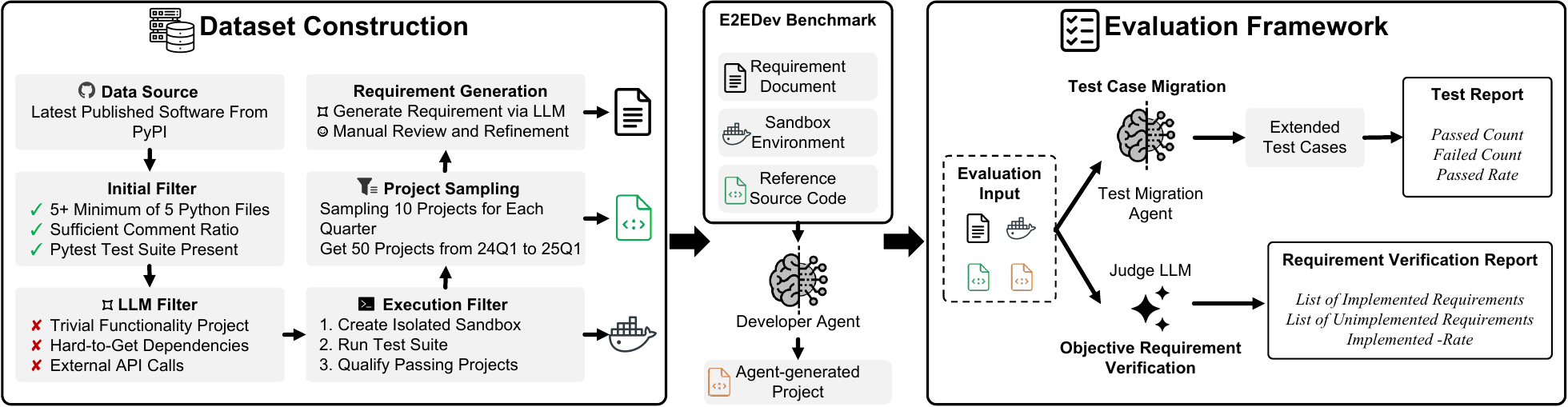}
    \caption{The dataset construction and evaluation workflow of \bench{}.}
    \label{fig:E2EDevBench_workflow}
    \Description[Workflow of \bench{} dataset construction and evaluation]{Workflow of \bench{} dataset construction and evaluation}
\end{figure*}

To bridge these gaps, we introduce the \textbf{\bench{}} (End-to-End Software Development Benchmark), constructed by curating 50 recent, high-quality projects from PyPI~\cite{PyPI} with an evaluation framework aimed at addressing the core difficulties in current assessment methodologies. The workflow for our dataset construction and the corresponding evaluation process is illustrated in Figure~\ref{fig:E2EDevBench_workflow}

\subsubsection{\textbf{Dataset Construction}}
\label{sec:dataset_construction}
Our dataset construction process is designed to select high-quality, real-world projects with reproducible environments. The process is divided into the following stages:
\begin{enumerate}[label=\arabic*., font=\bfseries, leftmargin=*, topsep=0.5em, itemsep=0.2em]
    \item \textbf{Source Data Collection and Initial Filtering:} We begin with a large-scale collection of projects from the PyPI~\cite{pypiBigQueryDatasets}. To ensure project quality, we apply a series of filtering rules, retaining only those projects that meet the following criteria: containing more than five Python source files, exhibiting a reasonable code-to-comment ratio, including a test suite, and utilizing \texttt{pytest}~\cite{pytestPytestDocumentation} as the testing framework.

    \item \textbf{LLM-based Filtering:} To ensure that projects are both meaningful and reliably executable within an isolated sandbox, we employ an LLM for a second filtering pass. This step systematically excludes projects with trivial functionality, those reliant on hard-to-get dependencies (e.g., requiring specific software or OS utilities), or those that make external API calls. This stage is critical for ensuring the reproducibility of the subsequent evaluation environment.

    \item \textbf{Execution-based Filtering:} For the projects that pass the filtering stages, we use automated scripts to attempt the construction of a runnable sandbox environment from their provided configuration files (e.g., \texttt{requirements.txt}). Subsequently, the script automatically executes the project's original test suite to confirm both the project's functional correctness and the completeness of the execution environment.
    
    \item \textbf{Sampling and Requirement Generation:} We employ a time-sliced sampling strategy. From the first quarter of 2024 to the first quarter of 2025, we randomly sample 10 standard-compliant projects each quarter, constituting an initial dataset of 50 projects. This strategy helps mitigate the risk of data leakage and allows the dataset to be continuously updated to reflect the latest development practices. 
    For each finalized project, we generate a requirements document by using an LLM to analyze the source code and README file and produce an initial draft based on a predefined template. This draft is then manually reviewed and revised to ensure accuracy and comprehensiveness.
    
\end{enumerate}

Ultimately, our construction process yielded a collection of 50 high-quality project development samples, with 10 projects sampled from each quarter between Q1 2024 and Q1 2025. Each instance in the dataset includes: 1) a manually verified requirements document that details the features to be implemented; 2) an interactive sandbox environment for the agent to operate in; 3) and the original source code and test cases, which serve as a reference for our evaluation framework.


\begin{table*}[]
\caption{The statistical comparison between existing repo-level generation benchmarks.}
\label{tab:benchmark_comparison}
\begin{adjustbox}{width=0.7\linewidth}
\begin{tabular}{|l|cccccc|} 
\hline
\textbf{Benchmark} & \multicolumn{1}{c}{\textbf{\# Tasks}} & \multicolumn{1}{c}{\textbf{\# Avg. Files}} & \multicolumn{1}{c}{\textbf{\# Avg. LOC}} & \multicolumn{1}{c}{\textbf{\#Avg. Tests}} & \textbf{Input Information} & \textbf{Data Leakage} \\ \hline\hline 
SoftwareDev~\cite{MetaGPT} & 70 & N/A\tnote{1} & N/A & N/A & Requirement & Potential \\ \hline 
ProjectDev~\cite{AgileCoder} & 14 & N/A & N/A & N/A & Requirement & Potential \\ \hline 
DevEval~\cite{DevEval(b)} & 22 & 4.4 & 377.8 & 10.18 & Req.\tnote{2} + Class Skeleton & Potential \\ \hline 
ProjectEval~\cite{ProjectEval} & 20 & 13.2 & 402.2 & 14.2 & \begin{tabular}[c]{@{}c@{}}Req. + Class Skeleton\\ /Function Skeleton\end{tabular} & Potential \\ \hline 
Commit0~\cite{commit0} & 54 & 205.19 & 24768.4 & 730.5 & Req. + Class Skeleton & Potential \\ \hline\hline 
\bench{} & 50 & 19.2 & 2011.5 & 119.7 & Requirement & Controlled \\ \hline 
\end{tabular}
\end{adjustbox}
\begin{tablenotes}
    \footnotesize
    \item[1] \textbf{N/A:} Not applicable, as these benchmarks do not have corresponding reference repositories. \textbf{Req.:} An abbreviation for Requirement. \textbf{Controlled:} Data leakage is can be eliminated by the experiment setting. \textbf{LOC:} Lines of Code.
\end{tablenotes}
\end{table*}

As detailed in Table~\ref{tab:benchmark_comparison}, \bench{} demonstrates a higher degree of complexity across multiple dimensions compared to existing repository-level benchmarks. In comparison to its counterparts, projects in \bench{} exhibit a significant increase in the average number of source files, lines of code (LoC), and test cases, and are accompanied by more detailed requirements documents. These characteristics indicate that \bench{} provides a more challenging and realistic testbed for evaluating the end-to-end development capabilities of autonomous agents.

\subsubsection{\textbf{Evaluation Framework}}
\label{sec:evaluation_framework}
As described in Section~\ref{sec:benchmark_background}, the evaluation of an agent tasked with developing a project from scratch based solely on a requirements document presents significant challenges. On one hand, the architectural implementation generated by the agent may differ substantially from that of the original project, rendering the original test suite directly unusable. On the other hand, a complete reliance on an ``LLM as Judge''~\cite{LLMasajudgeSurvey} for subjective scoring can yield unreliable results due to the inherent biases of the model. To address these challenges, we propose a hybrid evaluation methodology that combines automated test case migration with objective requirement verification.

\begin{enumerate}[label=\arabic*., font=\bfseries, leftmargin=*, topsep=0.5em, itemsep=0.2em]
    \item \textbf{Automated Test Case Migration:} After an agent completes the development process and submits its final project, we introduce an independent ``Test Migration Agent.'' The sole responsibility of this agent is to adapt the original project's test cases for use with the newly generated project. During this process, we employ technical constraints to strictly prevent the migration agent from modifying the code under evaluation. This approach enables us to leverage high-quality, original test cases for functional verification without providing them as input to the development agent, thereby granting the agent complete freedom in its design and implementation choices.
    \item \textbf{Objective Requirement Verification:} After obtaining the execution results from the migrated test cases, we proceed to the requirement verification stage. We provide an LLM with the test results, the agent-generated project code, and the original requirements document. Instead of soliciting a subjective score, we instruct the LLM to perform an objective marking task for each item on the requirements list, classifying it as either ``implemented'' or ``not implemented.'' Prior studies have showed that binary classification task to be more reliable than subjective scoring~\cite{ReproducibleResults,MTbench}. To further enhance the stability of our evaluation, we conduct three independent requests for the same assessment; a requirement is marked as definitively ``implemented or not implemented'' only if all three assessments concur. This process allows us to calculate a precise requirement completion rate as the primary evaluation metric, with the test pass rate serving as a supplementary indicator of functional correctness.
\end{enumerate}

From this hybrid process, we derive our key performance metrics. The functional quality is measured by the \textit{Pass-Rate}, calculated as the ratio of passed test cases (\textit{Passed Count}) to the total. The primary measure of success, however, is the \textit{Implementation Rate (Impl-Rate)}. This metric is computed from the requirement verification stage, representing the proportion of requirements unanimously marked as implemented (\textit{Impl-Req.}) relative to the total number of requirements. The total includes those unanimously marked as not implemented (\textit{Unimpl-Req.}) and those with inconclusive assessments across the three evaluation runs.

\subsection{Study Agent}
\label{sec:agent}
To systematically investigate the impact of different agent workflow designs on the performance of end-to-end software development tasks, we constructed a series of experimental agent systems. This section details our design philosophy and specific implementations.

Currently, software development agent systems can be broadly categorized into single-agent and multi-agent architectures. However, conducting a fair comparison between these two paradigms presents significant engineering challenges. As described in Section~\ref{sec:agent_background}, discrepancies in the underlying toolsets, environmental interaction capabilities, and implementation details across different research efforts often act as confounding variables, obscuring the intrinsic merits or demerits of the architectural designs themselves. For instance, in our preliminary explorations with existing multi-agent frameworks (e.g., ChatDev~\cite{ChatDev}, MetaGPT~\cite{MetaGPT}), we observed that their rigid toolchains and limited environmental interaction capabilities, such as an inability to flexibly invoke Bash commands for environment configuration, severely constrained their performance on complex tasks, leading to suboptimal quality in the generated code.

We posit that the core contribution of these frameworks lies not in their specific engineering implementations, but in their conceptualization of multi-agent task workflows, often based on established software development models like the waterfall model\cite{waterfall}. To isolate the influence of implementation-specific factors and facilitate a fair comparison under controlled conditions, we adopted a strategic approach: we selected the representative open-source project SWE-Agent~\cite{SWEagent} to serve as the unified foundational architecture for all our experimental agents. SWE-Agent provides a mature and flexible toolset, making it an ideal cornerstone for constructing complex agent systems~\cite{Leaderboard}. Building upon this foundation, we focused on implementing and comparing different task workflows, thereby ensuring that observed performance variations could be more directly attributed to the workflow designs themselves.
To maintain this focus, our primary goal was to investigate the effectiveness of structured task decomposition. We therefore simplified inter-agent collaboration, modeling it as a sequential handoff of tasks. Inspired by established software development paradigms like the waterfall model, we designed and implemented three representative agent workflow configurations to explore different divisions of responsibility:
\begin{enumerate}[label=\arabic*., font=\bfseries, leftmargin=*, topsep=0.5em, itemsep=0.2em]

\item \textbf{\agentd{} (Software Development Agent).}
In this configuration, all development tasks are handled by a single, all-purpose agent. This agent is instructed to follow a linear, self-iterative development process: it first reads the requirements document from the file system, then implements the required functionality in code, creates corresponding tests, and runs these tests. Finally, it reviews the test results and debugs the code, repeating this cycle until all tests pass.

\item \textbf{\agentdt{} (Developer-Tester).}
This configuration simulates a collaborative model with a separation of development and testing duties, utilizing two agents that operate sequentially. The first, a \textbf{Developer Agent}, focuses exclusively on development. It is responsible for reading the requirements document and completing the code implementation for all features, but it does not write or execute tests. The second, a \textbf{Tester Agent}, it receives the code completed by the Developer Agent, first comprehends the implementation, and then writes test cases based on the original requirements. Subsequently, it executes the tests, and performs any necessary debugging until all tests pass.

\item \textbf{\agentddt{} (Designer-Developer-Tester).}
This configuration simulating a more classical ``design-implement-verify'' waterfall workflow with three agents collaborating sequentially. At first, a \textbf{Designer Agent}, analyze the requirements document and generate a detailed project design document. Second, a \textbf{Developer Agent}, performs the concrete code implementation based on the design document produced by the Designer Agent and the original requirements. Third, a \textbf{Tester Agent}, conducting comprehensive testing and verification of the final implementation.
\end{enumerate}

\paragraph{\textbf{Experimental Setup:}}
All our experiments were powered by two leading LLMs Gemini-2.5-Pro and Gemini-2.5-Flash~\cite{Gemini25blog}. We selected these models for their state-of-the-art capabilities, which serve as a representative foundation for our study~\cite{gemini2.5}. As our research prioritizes the analysis of workflow design over a comparative LLM benchmark, and to maintain a feasible computational scope, we intentionally limited our experiments to these two models. In all trials, the generation temperature was set to 0.2, and a maximum of 200 agent steps was allocated for each task to ensure consistency and termination.
For the processes of benchmark construction and evaluation, we standardized on using Gemini-2.5-Pro to ensure maximum consistency and capability. Specifically, it was the exclusive model used for the dataset annotation, the ``Test Case Migration'', and the ``LLM as Judge'' protocol. Furthermore, the ``Test Case Migration'' itself was performed by a dedicated agent configured in the Single Agent setup, with the explicit task of migrating test cases.

\subsection{Research Questions}
This study investigates the following research questions:

\begin{enumerate}[label={\scalebox{2}{\textbullet}},leftmargin=*]
\item \textbf{RQ1: How effective is the requirement-driven ``LLM as Judge'' approach in \bench{}?}
In this RQ, we evaluate the effectiveness of this requirement-driven approach by comparing its assessments against both test suite outcomes and human expert judgments.
\item \textbf{RQ2: What are the end-to-end development capabilities of current software development agents on the \bench{}?}
In this RQ, we systematically assess agents with different architectures and base models, focusing on their task success rates, stability, and consistency across multiple trials.
\item \textbf{RQ3: What are the key challenges and root causes of agent failures in development tasks?}
In this RQ, we identify the root causes of failure by analyzing two aspects: identifying common patterns among the unfulfilled requirements, and their common failure modes during key development stages.
\end{enumerate}

\section{Study Results}

\subsection{RQ1: Effectiveness of Evaluation Framework}
\label{sec:rq1}
To validate the effectiveness of the requirement-driven LLM-as-a-judge evaluation framework in \bench{}, we conducted analyses on two key aspects: the comprehensiveness of the test cases and their consistency with human expert assessments.

First, we evaluated the effectiveness of the test case migration scheme, which underpins the comprehensiveness of the evaluation. We performed a statistical analysis of the test run results generated during the RQ2 experiments (detailed in Section~\ref{sec:rq2}). As shown in Table~\ref{tab:RQ1ComparativeAnalysisTable}, the test cases are divided into three categories: ``{Original Tests}'' (generated by the \texttt{SDAgent} during development), ``{Extended Tests}'' (generated by the \texttt{Test-Migration Agent}), and ``{All Tests}'' (the union of the first two). The table reports key metrics for execution outcomes (\textit{pass and failure counts, pass-rate}) and code coverage analysis (line coverage count, branch coverage count). The results indicate that, compared to the ``{Original Tests}''’ \textit{pass-rate} of 93.14\%, the ``{Extended Tests}'' exhibit a significantly lower \textit{pass-rate} of 59.66\%, demonstrating that the ``{Extended Tests}'' effectively uncover latent defects missed by the ``{Original Tests}''. Furthermore, the combined test suite (``All Tests'') significantly improves coverage: the average number of covered lines increases from 156.46 to 188.1 (a 20.22\% improvement), and covered branches increase from 39.83 to 50.5 (a 26.79\% improvement). These results confirm that our test case migration scheme enhances both error-detection capability and coverage breadth, thereby providing a more reliable basis for subsequent requirement-driven evaluations.


\begin{table}[]
\caption{Performance analysis of the Original Tests and Extended Tests.}
\label{tab:RQ1ComparativeAnalysisTable}
\begin{adjustbox}{width=\linewidth}
\begin{tabular}{|l|ccccc|}
\hline
\textbf{Tests Type} & \textbf{\textit{Avg. Passed}} & \textbf{\textit{Avg. Failed}} & \textbf{\textit{Pass-Rate}} & \textbf{\begin{tabular}[c]{@{}c@{}}\textit{Avg. Covered}\\ \textit{Lines}\end{tabular}} & \textbf{\begin{tabular}[c]{@{}c@{}}\textit{Avg. Covered}\\ \textit{Branches}\end{tabular}} \\ \hline\hline
\textbf{Original Tests}      & 13.84                   & 1.02                    & 93.14\%              & 156.46                         & 39.83                             \\ \hline
\textbf{Extended Tests}      & 15.07                   & 10.19                   & 59.66\%              & 142.17                         & 32.01                             \\ \hline
\textbf{All Tests}           & 28.89                   & 11.23                   & 72.01\%              & 188.1                          & 50.5                              \\ \hline
\end{tabular}
\end{adjustbox}
\end{table}

\begin{figure}
    \centering
    \includegraphics[width=1\linewidth]{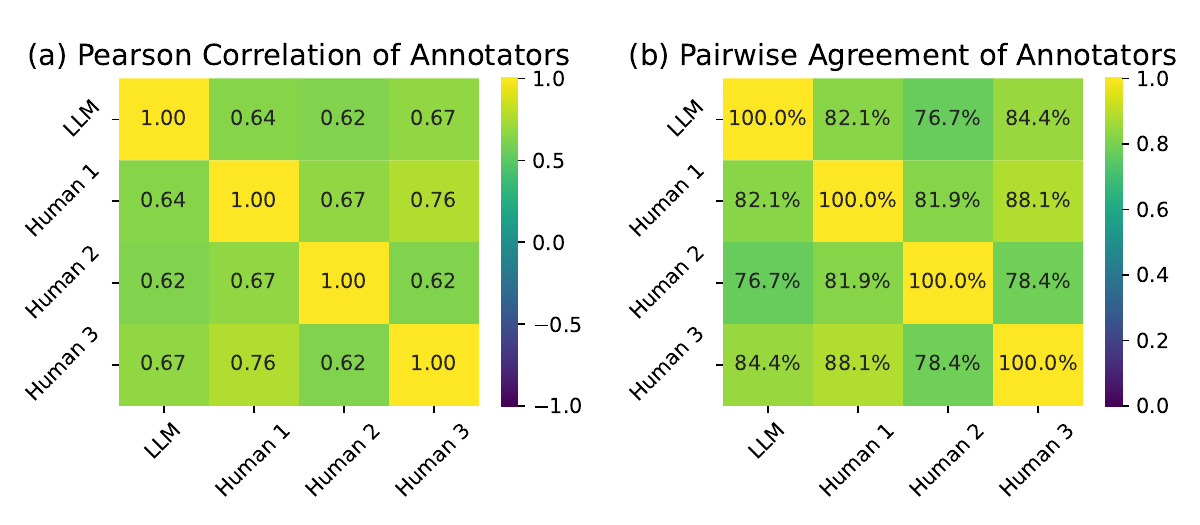}
    \caption{Analysis of agreement between LLM-as-Judge and human experts.}
    \label{fig:annotator_analysis_combined}
\end{figure}

Second, we assessed the consistency between the requirement-driven LLM-as-judge and human expert evaluations. We randomly selected 10 agent-generated projects with 546 requirements and invited three computer science graduate students as human experts. Both the LLM (Gemini-2.5-Pro) and the human experts independently annotated whether each requirement had been successfully implemented by the agent. We employed Pearson’s correlation coefficient~\cite{Pearsoncorrelation} and pairwise agreement rate~\cite{MTbench} to measure annotation agreement, with results presented in Figure~\ref{annotator_analysis_combined}. The findings show a high level of agreement between the LLM and all three experts. Notably, the Pearson's correlation coefficients were all above 0.62, indicating a strong positive correlation between the LLM's assessments and those of the human experts. Furthermore, the pairwise agreement rate between the LLM and the experts ranged from 76\% to 84\%, a level comparable to the inter-expert agreement rate of 78\% to 88\%. This collectively demonstrates that the requirement-driven LLM-as-a-judge evaluation method possesses high reliability and validity.

\findingbox{Conclusion 1:}{The test-migration and requirement-driven LLM-as-a-judge evaluation method is effective, it not only significantly increases test coverage and uncovers additional defects by generating new test cases, but also exhibits a high degree of agreement with human expert evaluations, thereby demonstrating the reliability of this evaluation metric.}

\subsection{RQ2: Performance of SDAgents under \bench{}}
\label{sec:rq2}

\begin{table*}[]
\caption{Performance of different Agent–LLM combinations on \bench{}.}
\label{tab:RQ2AgentPerformance}
\begin{adjustbox}{width=0.85\linewidth}
\begin{tabular}{ccccccccccccc}
\hline
\multicolumn{1}{|c|}{\textbf{LLM}} & \textbf{\textit{Impl-Req.}} & \textbf{\textit{Unimpl-Req.}} & \multicolumn{1}{c|}{\textbf{\textit{Impl-Rate.}}} & \textbf{\textit{Passed}} & \textbf{\textit{Failed}} & \multicolumn{1}{c|}{\textbf{\begin{tabular}[c]{@{}c@{}}\textit{Passed}\\ \textit{Rate}\end{tabular}}} & \textbf{\begin{tabular}[c]{@{}c@{}}\textit{Tokens}\\ \textit{Sent (M)}\end{tabular}} & \textbf{\begin{tabular}[c]{@{}c@{}}\textit{Token}\\ \textit{Gen (K)}\end{tabular}} & \multicolumn{1}{c|}{\textbf{\textit{Cost (\$)}}} & \textbf{\begin{tabular}[c]{@{}c@{}}\textit{Code}\\ \textit{Lines}\end{tabular}} & \textbf{\begin{tabular}[c]{@{}c@{}}\textit{Code}\\ \textit{Files}\end{tabular}} & \multicolumn{1}{c|}{\textbf{\textit{Steps}}} \\ \hline
\multicolumn{13}{c}{\textbf{\texttt{SDAgent-Single}}} \\ \hline
\multicolumn{1}{|c|}{\textbf{Flash}}    & 27.14              & 23.32                & \multicolumn{1}{c|}{48.46\%}            & \textbf{42.35}                & 12.96                & \multicolumn{1}{c|}{76.57\%}              & 20.90                         & 161.24                      & \multicolumn{1}{c|}{6.67}                    & 533.10                 & 4.22                    & \multicolumn{1}{c|}{152.74}             \\ 
\multicolumn{1}{|c|}{\textbf{Pro}}      & 25.14              & 21.86                & \multicolumn{1}{c|}{42.97\%}            & 28.58                & 8.92                 & \multicolumn{1}{c|}{76.21\%}              & 4.96                          & 28.45                       & \multicolumn{1}{c|}{6.48}                    & 209.02                 & 3.50                    & \multicolumn{1}{c|}{80.12}              \\ 
\multicolumn{1}{|c|}{\textbf{Mean}}     & 26.14              & 22.59                & \multicolumn{1}{c|}{45.72\%}            & 35.47                & 10.94                & \multicolumn{1}{c|}{76.39\%}              & 12.93                         & 94.85                       & \multicolumn{1}{c|}{6.58}                    & 371.06                 & 3.86                    & \multicolumn{1}{c|}{116.43}             \\ \hline
\multicolumn{13}{c}{\textbf{\texttt{SDAgent-DT}}} \\ \hline
\multicolumn{1}{|c|}{\textbf{Flash}}    & 26.60              & 19.30                & \multicolumn{1}{c|}{45.47\%}            & 28.04                & 9.16                 & \multicolumn{1}{c|}{75.38\%}              & 19.14                         & 192.37                      & \multicolumn{1}{c|}{6.22}                    & 307.96                 & 4.42                    & \multicolumn{1}{c|}{90.60}              \\ 
\multicolumn{1}{|c|}{\textbf{Pro}}      & \textbf{31.30}              & \textbf{16.14}                & \multicolumn{1}{c|}{\textbf{53.50\%}}            & 31.34                & \textbf{7.86}                 & \multicolumn{1}{c|}{\textbf{79.95\%}}              & 5.24                          & 49.65                       & \multicolumn{1}{c|}{7.05}                    & 328.28                 & 6.58                    & \multicolumn{1}{c|}{77.36}              \\ 
\multicolumn{1}{|c|}{\textbf{Mean}}     & 28.95              & 17.72                & \multicolumn{1}{c|}{49.48\%}            & 29.69                & 8.51                 & \multicolumn{1}{c|}{77.66\%}              & 12.19                         & 121.01                      & \multicolumn{1}{c|}{6.64}                    & 318.12                 & 5.50                    & \multicolumn{1}{c|}{83.98}              \\ \hline
\multicolumn{13}{c}{\textbf{\texttt{SDAgent-DDT}}} \\ \hline
\multicolumn{1}{|c|}{\textbf{Flash}}    & 12.82              & 31.88                & \multicolumn{1}{c|}{22.63\%}            & 25.57                & 12.49                & \multicolumn{1}{c|}{67.18\%}              & 18.27                         & 151.64                      & \multicolumn{1}{c|}{5.86}                    & 340.86                 & 4.30                    & \multicolumn{1}{c|}{108.06}             \\ 
\multicolumn{1}{|c|}{\textbf{Pro}}      & 19.18              & 27.56                & \multicolumn{1}{c|}{32.79\%}            & 26.74                & 15.52                & \multicolumn{1}{c|}{63.27\%}              & 3.60                          & 34.62                       & \multicolumn{1}{c|}{4.85}                    & 344.92                 & 13.74                   & \multicolumn{1}{c|}{86.16}              \\ 
\multicolumn{1}{|c|}{\textbf{Mean}}     & 16.00              & 29.72                & \multicolumn{1}{c|}{27.71\%}            & 26.16                & 14.00                & \multicolumn{1}{c|}{65.23\%}              & 10.94                         & 93.13                       & \multicolumn{1}{c|}{5.36}                    & 342.89                 & 9.02                    & \multicolumn{1}{c|}{97.11}              \\ \hline
\end{tabular}
\end{adjustbox}
\end{table*}

Building upon the established validity of the \bench{} framework, we conducted a comprehensive performance evaluation. Specifically, three agent workflow configurations (\agentd{}, \agentdt{}, and \agentddt{}) were each coupled with two model variants (Gemini-2.5-Pro and Gemini-2.5-Flash), with detailed results presented in Table~\ref{tab:RQ2AgentPerformance}. The table shows the average performance of all instances across three key metric categories: requirement completion, test execution performance, and resource-efficiency overhead.

The experimental findings reveal that the \agentdt{} paired with the Gemini-2.5-Pro model achieved the highest requirement implementation rate of 53.50\%. This demonstrates that, when tasked with realistic end-to-end development challenges, the agent can satisfy approximately half of the functional requirements. However, this success is tempered by a significant challenge: the other half of the requirements remained unfulfilled, exposing reliability bottlenecks in the agents’ autonomous development processes. This suggests that, in executing complex tasks independently, current agents lack sufficient robustness and problem-solving capabilities to fully supplant human developers, which is a critical obstacle for future research.

Moreover, agent operation incurs substantial resource overhead. For instance, the top-performing \agentdt{} + Pro configuration consumed on average 5.24 million \textit{TokensSent} and 49.65 thousand \textit{TokenGen} per project, with an estimated cost of 7.05\$. Such high overhead underscores the urgent need to optimize agent workflows and enhance their resource-utilization efficiency.

\findingbox{Conclusion 2:}{Current advanced software development agents possess the capability to fulfill partial development requirements; nevertheless, their overall performance and cost-efficiency present major areas for improvement.}

\begin{figure}
    \centering
    \includegraphics[width=1\linewidth]{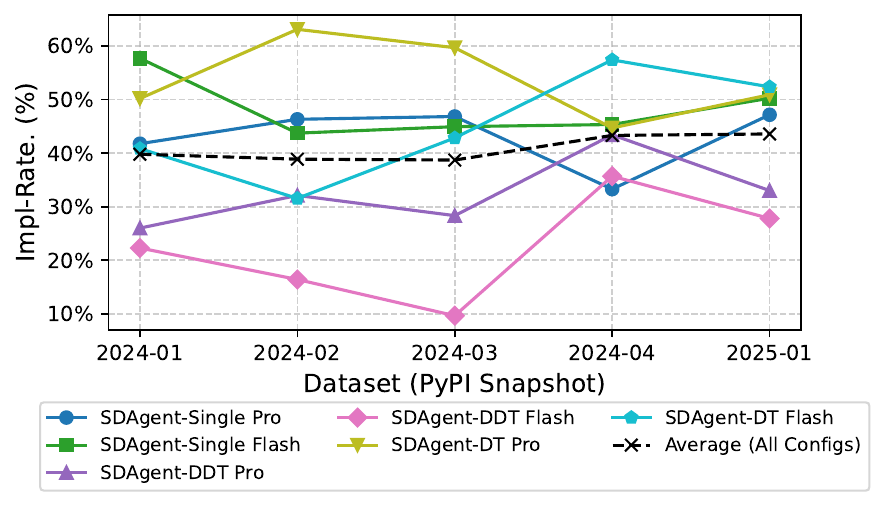}
    \caption{Agent performance stability across different time periods.}
    \label{fig:ImplementationRatiobyTime}
\end{figure}

Although the data in Table~\ref{tab:RQ2AgentPerformance} demonstrate the agents’ remarkable capabilities, it is crucial to verify the authenticity of these results and exclude any potential data leakage. To this end, we assessed the agents’ generalization ability using temporally unseen datasets. Specifically, \bench{} can be partitioned by project release date, including open-source projects released in 2025-Q1. Given that the Gemini series models employed in our experiments have a knowledge cutoff of January 2025~\cite{geminimodelcard}, this dataset can be regarded as completely unseen. Figure\ref{fig:ImplementationRatiobyTime} depicts the \textit{Impl-Rate.} across five time intervals spanning 2024 and 2025-Q1 for different agent-model combinations. The results reveal significant performance stability: when processing the unseen 2025-Q1 data, the agents did not exhibit the anticipated performance drop that might arise from data leakage. Specifically, the average requirement implementation rate in 2025-Q1 (42\%) is virtually on par with the 2024 average level (40\%). This temporal consistency strongly indicates that the agents’ performance reflects genuine generalization capability rather than memorization of the training data.

\findingbox{Conclusion 3:}{The agent’s performance remained stable on the unseen dataset, thereby excluding the possibility of data leakage and demonstrating its genuine generalization capability in development tasks.}

\begin{figure}[htbp]
    \centering
    \includegraphics[width=1\linewidth]{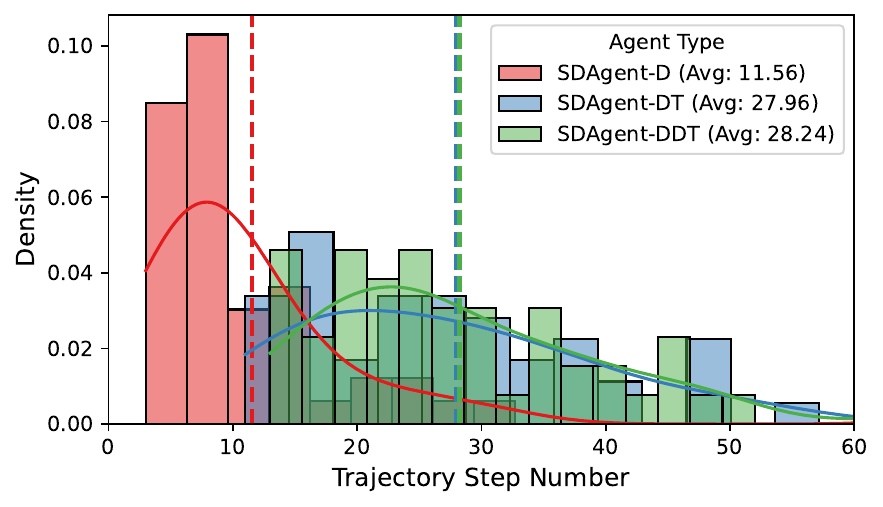}
    \caption{Distribution of first test writing step by agent.}
    \label{fig:first_test_step_distribution}
\end{figure}

In our comparative analysis of agent architectures, we found that workflow design is a key determinant of performance. As shown in Table~\ref{tab:RQ2AgentPerformance}, the \agentd{} achieves the highest requirement implementation rate (49.48\%), outperforming \agentd{} (45.72\%) and far exceeding the lowest-performing \agentddt{} (27.71\%). To probe the roots of this performance gap, we analyzed each agent’s behavioral patterns, especially the coordination between development and testing. Figure~\ref{fig:first_test_step_distribution} depicts the distribution of the step at which each agent first writes a test case. The data indicate that, due to its required develop-then-test workflow, the \agentdt{} begins writing tests significantly later, on average at step 27.96. In contrast, the \agentd{} initiates testing much earlier, at step 11.56 on average, reflecting a highly interleaved code-and-test strategy. We posit that this seemingly agile iterative mode is the primary cause of \agentd{}’s suboptimal performance. Frequent context switches between coding and debugging substantially increase the risk of context loss. When errors are detected during testing, the agent tends to fall into a reactive local-fixing loop, deviating from the original global development plan. This ``subtask immersion'' ultimately leads to neglect of higher-level objectives and failure to satisfy many core requirements. By contrast, the more structured workflow of the \agentdt{} effectively avoids this trap.

Furthermore, we found that although the \agentddt{} adopts a similar develop-then-test strategy to the top-performing \agentdt{}, its requirement implementation rate remains the lowest. Given that its \textit{Dev} and \textit{Test Agent} are configured identically to those of the \agentd{}, we hypothesize that the performance bottleneck originates in its \textit{Design Agent}. We attribute this to the fact that the planning blueprint produced by the \textit{Design Agent} negatively impacts subsequent development. Once the \textit{Dev Agent} receives what it perceives as an authoritative implementation plan, it tends to prioritize this plan over direct engagement with the original requirement document, even though the latter is provided as context. If the design plan itself is flawed, such errors are faithfully propagated downstream, resulting in a final product that diverges from the actual requirements. In contrast, the \textit{Dev Agent} within the \agentdt{}, without a predefined plan, is compelled to interpret and implement directly from the original requirement document. This direct interaction ensures that the development process more faithfully adheres to user needs, thereby achieving a higher requirement implementation rate.

\findingbox{Conclusion 4:}{The superior performance of the Dev‑Test workflow demonstrates that different workflow designs and agent orchestrations can greatly affect overall performance. A single‑agent approach is not necessarily optimal. A well‑structured division of tasks and agents can effectively reduce task complexity, whereas an inappropriate decomposition can substantially degrade agent performance.}

When comparing the performance of LLMs of different scales, we made a notable observation: on the core metric of requirement fulfillment rate, the relatively weaker Flash model did not exhibit the expected significant performance gap with the more powerful Pro model. For instance, in the \agentd{}, Flash achieved a fulfillment rate of 48.46\%, even outperforming Pro's 42.97\%.
We hypothesize that this can be attributed to two factors. First, the Flash model's performance is achieved by incurring significantly higher computational costs. Data in Table~\ref{tab:RQ2AgentPerformance} show that its overhead in \textit{TokensSent} is multiple times that of the Pro model (e.g., 20.90M for Flash vs. 4.96M for Pro in the \agentd{}). This suggests that Flash model compensates for its weaker single-turn reasoning capabilities through more interaction rounds and trial-and-error. This lengthy and costly exploration is effective sometimes, as each iteration may partially or coincidentally fulfill certain requirements. Consequently, Flash's high fulfillment rate is the cumulative result of successes aggregated through extensive trial-and-error. Second, the Pro model is more efficient in execution (requiring fewer tokens on average), suggesting that the Pro model may be more prone to prematurely triggering the agent's self-submission condition, judging the task as complete before all requirements are truly met. This pattern of ``premature termination'' hinders the full exploitation of its advanced capabilities.
\findingbox{Conclusion 5:}{Weaker models can match the performance of more capable models to some extent by incurring higher computational overhead. Meanwhile, for more powerful models, the current simplistic agent self-submission mechanism may be constraining their full potential.}

\begin{figure}[htbp]
    \Description{A chart of three independent runs}
    \centering
    \includegraphics[width=1\linewidth]{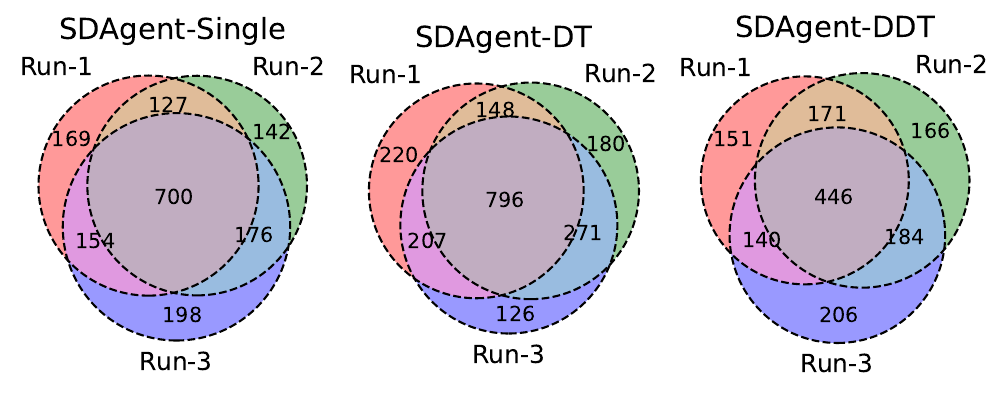}
    \caption{Implemented requirements comparison across 3 runs for each agent.}
    \label{fig:3Runs}
\end{figure}

In addition, given the inherent stochasticity of LLMs, a common practice is to improve outcomes via ``best-of-N'' sampling or result aggregation~\cite{bestofnalphacode,bestofnMajorityVoting}. We therefore evaluated the impact of this randomness on agent performance through three independent runs (N=3, using Gemini-2.5-Pro). To this end, a Venn diagram illustrating the overlap of successful implementations is presented in Figure~\ref{fig:3Runs}.

The Venn diagram reveals a high degree of consistency in the agents' performance. Taking the best-performing \agentdt{} as an example, a large core set of 796 requirements was jointly fulfilled by all three runs. However, we also observed that a substantial number of requirements (526) were uniquely fulfilled by only one of the three runs. This indicates that a "best-of-N" strategy is a feasible path to improving overall performance by capturing these unique successes. Nevertheless, given the high computational cost of a full agent run, the cost-effectiveness of such an approach becomes a critical concern that requires careful optimization.

\findingbox{Conclusion 6:}{While an agent's performance shows high core consistency, multiple runs can capture unique requirement fulfillments, indicating potential for "best-of-N" strategies. However, this approach faces a significant challenge in its high computational cost, making the development of cost-effective aggregation methods a key research direction.}

\begin{table*}[htbp]
    \centering
    \caption{A taxonomy of unimplemented requirements and their distribution.}
    \label{tab:failure_classification}
    \begin{tabularx}{\linewidth}{@{} |p{6cm} |X| r @{}} 
        \hline
        \textbf{Category (Distribution)} & \textbf{Description} \\
        \hline\hline
        \textbf{1. Missing Component or Feature (32.4\%)} & 
        The required functionality is entirely absent from the codebase. \\
        \hline\hline
        
        \textbf{2. Incomplete Implementation (18.1\%)} & 
        The feature is partially present but fails to meet the full scope of the requirement. \\ 
        
        2.1 Placeholder or Stub Implementation (6.4\%) & 
        A structural element is defined, but its implementation is empty or non-functional. \\
        
        2.2 Incomplete Logic or Behavior (11.7\%) & 
        The implementation handles the primary use case but is missing required sub-features or edge case handling. \\
        \hline\hline
        \textbf{3. Incorrect Implementation (34.1\%)} & 
        The feature has been implemented in its entirety, but it behaves incorrectly or produces the wrong results.  \\ 
        
        3.1 Flawed Core Logic (17.7\%) & 
        The fundamental algorithm or business logic used to implement the feature is incorrect. \\
        
        3.2 Incorrect Output or Data Formatting (6.0\%) & 
        The final output does not conform to the specified format. (e.g., incorrect string formatting, data structure, or file content). \\
        
        3.3 Incorrect API or Signature (10.4\%) & 
        The implementation violates the required software architecture or public-facing contracts, such as incorrect class inheritance. \\
        \hline\hline
        \textbf{4. Dependent or Upstream Failure (15.3\%)} & 
        This is a cascading failure where the component fails due to its reliance on a defective or unimplemented upstream dependency. \\
        \hline
    \end{tabularx}
\end{table*}

\subsection{RQ3: Study of Failure Modes}
To investigate the key challenges and root causes of agent task failures, we conducted an in-depth error analysis to understand both the types of unimplemented requirements that occurred and their underlying root causes.

To start with, drawing upon prior work~\cite{whydo}, we established an error taxonomy developed based on Grounded Theory~\cite{GroundedTheory} through an iterative manual analysis of unimplemented requirements. The full taxonomy is presented in Table~\ref{tab:failure_classification}. Specifically, we classify unimplemented requirements into four high-level categories: Missing Component or Feature, where functionality is entirely absent; Incomplete Implementation, where a feature is only partially realized; Incorrect Implementation, where the logic is flawed; and Dependent or Upstream Failure, where the error stems from other system parts. 

For a large-scale quantitative analysis, we randomly sampled 1,000 unimplemented requirements from 50 projects in the Pro model experiments across three agents' configurations. We employed an ``LLM pre-annotation, then human refine'' methodology to classify these samples, ensuring the accuracy and reliability of the annotations.
Table~\ref{tab:failure_classification} also illustrates the distribution of each unimplemented requirement type. The analysis reveals a key pattern: failures stem more from ``omission and validation failure'' than from ``implementation errors.'' The largest category of failures was due to components or features being completely omitted (32.4\%), followed by requirements being only partially implemented (21.7\%). Both of these error types can be attributed to the LLM's inadequate comprehension of the requirements. Furthermore, 15.3\% of failures were the result of cascading failures from upstream tasks. In contrast, genuine implementation errors accounted for 34.1\% of the failures. Aggregating these categories reveals that the vast majority of failures are not implementation errors at the code level, but rather omissions during the planning and execution process.


\findingbox{Conclusion 7:}{The primary failure mode for agents is ``omission and validation failure,'' not ``implementation error.'' Most requirement failures stem from requirements being completely overlooked, partially implemented, or blocked by upstream errors. 
}

Next, we traced the root causes of failure from the perspective of agent capabilities. Employing the same annotation method, we attributed failures to different stages of the agent's execution trace. Using the same annotation method, we classified failures into three distinct categories based on their origin, as presented in Table~\ref{tab:failure_root_cause_compact}. These categories are: \textbf{Task Planning}, where errors occur before implementation begins, such as omitting or misinterpreting requirements; \textbf{Task Execution}, where failures happen during the development process, e.g., technical roadblocks, context forgetting~\cite{inftyBench,LV-Eval}; and \textbf{Task Verification}, where the agent fails to properly assess its own work, (e.g., creating insufficient test cases and premature submission).

The results in Table~\ref{tab:failure_root_cause_compact}, clearly indicate that ``Task Planning'' is the primary bottleneck in current agent systems, accounting for 55.8\% of all issues. Specifically, the most prevalent problems are \textit{requirements being directly overlooked} (27.9\%) or \textit{misunderstood and oversimplified} (22.2\%) during the planning phase. In stark contrast, direct failures genuinely caused by insufficient core LLM capabilities (e.g., the inability to solve a specific technical challenge) account for only a minimal fraction (5.6\%). However, these few core capability failures have a disproportionately large downstream impact, potentially causing the subsequent 21\% of requirements to be blocked or invalidated due to dependencies. This reveals that in a complex task chain, the failure of a few critical nodes can trigger a widespread cascading effect.

\begin{table*}[htbp]
    \centering
    \caption{A taxonomy of failure root causes and their distribution.}
    \label{tab:failure_root_cause_compact}
    \begin{tabularx}{\linewidth}{@{} |p{5.3cm} |X| @{}} 
    \hline
        \textbf{Category (Distribution)} & \textbf{Description} \\
        \hline\hline
        \textbf{1. Task Planning (55.8\%)} & \\ 
        \textbf{1.1} Requirement Omission (27.9\%) & The agent entirely overlooks a requirement, making no attempt to implement it. \\
        
        \textbf{1.2} Requirement Misinterpretation (22.2\%) & The agent's understanding of the requirement is flawed (e.g., misinterpreting logic, specs, or constraints), leading to an implementation that fails to meet the specification. \\
        
        \textbf{1.3} Architectural Design Flaw (5.6\%) & A flawed architectural decision made early on renders subsequent tasks impossible to implement correctly. \\
        \hline\hline
        \textbf{2. Task Execution (38.6\%)} & \\ 
        \textbf{2.1} Technical Capability Bottleneck (3.2\%) & The agent fails because it cannot overcome the task's inherent technical complexity. \\
        
        \textbf{2.2} Context \& Task Forgetting (1.9\%) & The agent loses track of key information or planned tasks over long interactions, leading to omitted actions or a logically broken implementation. \\
        
        \textbf{2.3} Superficial Implementation (12.5\%) & The agent implements only the core 'happy path' of a requirement, ignoring necessary edge cases, error handling, and robustness requirements. \\
        \textbf{2.4} Dependent Feature Failure (21.0\%) & A cascading failure caused by the incorrect implementation of an upstream feature it depends on. \\
        \hline\hline
        
        \textbf{3. Task Verification (5.7\%)} & \\ 
        \textbf{3.1} Inadequate Verification (5.7\%) & The agent's own tests are too simplistic or incomplete to detect flaws in its implementation. \\
        
        \hline
    \end{tabularx}
\end{table*}

\findingbox{Conclusion 8:}{The core constraint of current agent systems is ``planning and comprehension''. The vast majority of failures stem from the omission or misinterpretation of requirements during the planning phase. This suggests that enhancing an agent's ability to comprehend, decompose, and validate requirements is a more critical research direction.}

\section{Implications and Discussions}
Our empirical study yields several important implications for the future of LLM-based software development agents.

\parabf{Requirement-driven evaluation is effective.}
Our study demonstrates that the proposed hybrid methodology, combining test-case migration with a requirement-driven ``LLM as Judge,'' proves to be highly effective. It not only increases test coverage to uncover hidden defects but also shows a strong correlation with human expert judgment (Conclusion 1). This implies that future research should adopt more holistic evaluation frameworks that measure not just functional correctness but also the granular fulfillment of user requirements, thereby providing a more reliable measure of an agent's true capabilities.

\parabf{Autonomous software development is a promising reality.}
Our findings confirm that end-to-end software development by LLM-based agents is feasible. State-of-the-art agents demonstrate a genuine ability to generalize and complete a substantial portion of complex, real-world development tasks on unseen data (Conclusion 2 \& 3). This provides strong evidence that this research direction holds significant promise. However, the path to full autonomy is still fraught with challenges, particularly concerning the reliability and computational cost of these systems. This suggests that while the long-term vision is achievable, near-term research should focus on enhancing robustness and efficiency to make these agents practical for real-world deployment.

\parabf{Agent orchestration is a critical performance factor.}
This research highlights that agent architecture is important. A well-designed multi-agent workflow, such as the Developer-Tester configuration, can significantly reduce task complexity and outperform a single-agent approach (Conclusion 4), while poor design can hamstring even the most powerful models. This finding implies that future work should explore and optimize agent collaboration patterns and task decomposition strategies as an important direction for improving performance.

\parabf{The core bottleneck is comprehension.}
A key finding from our analysis is that the principal bottleneck for agents is planning and comprehension. The vast majority of failures stem not from an inability to write correct code, but from the omission, misinterpretation, or inadequate verification of requirements (Conclusion 7 \& 8). This indicates that the core challenges are cognitive—related to planning and understanding—rather than technical execution. This has profound implications for the research community: a greater emphasis must be placed on enhancing an agent's ability to comprehend specifications, construct complete plans, and perform rigorous self-validation. Improving the agent's ``requirement engineering'' skills may yield greater returns than simply improving its code generation proficiency.

\section{Threats to Validity}
\label{sec:threats}

\noindent\textbf{Internal Validity.} The main threat arises from potential implementation errors in our unified agent framework, which could create confounding variables in our architectural comparison. We mitigated this by performing careful code reviews and making our entire experimental setup publicly available for scrutiny and replication~\cite{E2EDevStudy}.

\noindent\textbf{External Validity.} This threat arises from our choice of LLMs and the quality of our benchmark. To mitigate it, we selected two SOTA models, Gemini-2.5-Pro and -Flash. As our research focuses on the impact of agent architecture rather than a comparison of LLM capabilities, we contend these models are representative. Next, to ensure the benchmark's quality, all task requirements were rigorously revised by human experts, guaranteeing their quality.

\noindent\textbf{Construct Validity.} A significant threat is the reliability of our ``LLM as Judge'' component for evaluating requirement fulfillment. To address this, we validated our automated evaluation metric by demonstrating its high consistency with assessments performed by human experts, thereby ensuring that our framework accurately measures an agent's ability to meet user needs.

\section{Conclusion}
In this paper, we conducted a systematic evaluation of LLM-based agents on a new, challenging end-to-end software development benchmark. Our study, facilitated by a hybrid evaluation framework and controlled agent implementations, reveals that while current agents can fulfill approximately half of the requirements (\~ 50\%), their success is critically determined by their architectural design. The primary bottleneck is not in the execution phase but rather failures during the planning and verification phases, such as the omission or misinterpretation of requirements. These findings strongly suggest that future research must prioritize enhancing agents' capabilities in requirement comprehension and self-assessment. Our work provides a benchmark and framework to guide and measure progress in this vital direction.

\bibliographystyle{ACM-Reference-Format}
\bibliography{ref}

\end{document}